  \documentclass[10pt,aps,pra,reprint,showpacs,superscriptaddress]{revtex4-1}
\usepackage{dcolumn}
\usepackage[english]{babel}
\usepackage[T1]{fontenc}
\usepackage{natbib}
\usepackage{soul}
\usepackage{amssymb}
\usepackage{bm}
\usepackage{amsmath}
\usepackage[mathcal]{eucal}
\usepackage{color}
\usepackage{graphicx}
\usepackage{hyperref}
\usepackage{mathtools}
\usepackage{url}
\usepackage{tikz}
\usetikzlibrary{shapes}
\usetikzlibrary{decorations}
\usetikzlibrary{arrows}
\usetikzlibrary{matrix}

\newcommand {\nc} {\newcommand}
\nc {\beq} {\begin{eqnarray}}
\nc {\eeqn} [1] {\label{#1} \end{eqnarray}}
\nc {\eoln} [1] {\label{#1} \\}
\nc {\eol} {\nonumber \\}
\nc {\rref} [1] {(\ref{#1})}
\nc {\Eq} [1] {Eq.~(\ref{#1})}
\nc {\Ref} [1] {Ref.~\cite{#1}}
\nc {\la} {\mbox{$\langle$}}
\nc {\ra} {\mbox{$\rangle$}}
\nc {\dem} {\mbox{$\frac{1}{2}$}}
\nc {\cP} {\mathcal{P}}
\nc {\cN} {\mathcal{N}}
\nc {\ve} [1] {\mbox{\boldmath $#1$}}
\nc {\arrow} [2] {\mbox{$\mathop{\rightarrow}\limits_{#1 \rightarrow #2}$}}
\nc {\red}[1] {\textcolor{red}{#1}}
\nc {\mc}[3] {\multicolumn{#1}{#2}{#3}}
\nc {\dd}{\, \mathrm{d}}
\nc {\bs}[1]{\boldsymbol{#1}}
\nc {\ket}[1]{\vert #1 \rangle}
\nc {\bra}[1]{\langle #1 \vert}
\nc {\abs}[1]{\vert #1 \vert}
\nc {\avg}[1]{\langle #1 \rangle}
\nc {\braket}[2]{\langle #1 \vphantom{#2} \vert #2 \vphantom{#1} \rangle}
\nc {\abss}[1]{\left| #1 \right|}

\begin{document}

\title{\textit{Ab initio} calculations of hyperfine structures of zinc and evaluation of the nuclear quadrupole moment $Q(^{67}{\rm Zn})$}
\author{Jacek Biero{\'n}}
\email[]{Jacek.Bieron@uj.edu.pl}
\affiliation{Instytut Fizyki imienia~Mariana~Smoluchowskiego,
             Uniwersytet Jagiello{\'n}ski,
             ul.~prof.~Stanis{\l}awa {\L}ojasiewicza 11,
             Krak{\'o}w, Poland}
\author{Livio Filippin}
\affiliation{Chimie Quantique et Photophysique,
             Universit{\'e} libre de Bruxelles, B-1050, Brussels, Belgium}
\author{Gediminas Gaigalas}
\affiliation{Vilnius University,
             Institute of Theoretical Physics and Astronomy,
             Saul\.{e}tekio~av.~3, LT-10222, Vilnius, Lithuania}
%
\author{Michel~Godefroid}
\affiliation{Chimie Quantique et Photophysique,
             Universit{\'e} libre de Bruxelles, B-1050, Brussels, Belgium}
\author{Per J{\"o}nsson}
\affiliation{Department of Materials Science and Applied Mathematics,
             Malm\"o University, S-20506, Malm\"o, Sweden}
\author{Pekka Pyykk\"o}
\affiliation{Department of Chemistry, University of Helsinki, PO~Box~55
             (A.~I.~Virtasen aukio~1), FIN-00014 Helsinki, Finland}

\date{\today}

%
\begin{abstract}
The relativistic multiconfiguration Dirac-Hartree-Fock (MCDHF)
and the non-relativistic multiconfiguration Hartree-Fock (MCHF)
methods have been employed
to calculate the magnetic dipole and electric quadrupole
hyperfine structure constants of zinc.
The calculated electric field gradients
for the $ 4s 4p \,\, ^3 \! P^o_{1}$ and $ 4s 4p \,\, ^3 \! P^o_{2}$ 
states, together with experimental values of the electric quadrupole
hyperfine structure constants,
made it possible to extract a nuclear electric quadrupole moment
$Q(^{67}{\rm Zn}) = 0.122(10)$~{b}.
{The error bar has been evaluated in a quasi-statistical approach ---
the calculations had been carried out with eleven different methods,
and then the error bar has been estimated from the differences between
the results obtained with those methods.}
\end{abstract}

%
%

\maketitle

\section{Introduction}
\label{section.Introduction}

One of the {most accurate}
methods to determine nuclear quadrupole moments, $Q$,
is to combine
measured nuclear quadrupole coupling constants,
$B = e^2Qq/{(4\pi \epsilon_0)h}$ {(in frequency units)}, with
calculated or deduced electric field gradients ({EFG}),
$q$~\cite{Pyykko:2001}.
The aim of the present work is to apply {this} method to determine
improved nuclear quadrupole moments of zinc, the second 
most abundant {essential} trace element {in the human body, after iron \cite{Nie:2013a}}. 
%
%
For zinc, the standard value cited in the 2008 review of
Pyykk\"o~\cite{Pyykko:08bb}
and the 2016 review of
Stone~\cite{NJStone:2016} is still the 1969 value 
{obtained} by
Laulainen and McDermott~\cite{LaulainenMcDermott:1969}:
$Q(^{67}{\rm Zn})$ = 0.150(15)~{b} {(1~barn = 1~b =
$10^{-28} \, \mbox{m}^2$)}.
This value is based on the experimental
$B_{63}(^3 \! P^o_1) = -34.46(3)$~MHz for the
$4s4p \, ^3 \! P^o_1$ state and a $q$ value deduced from the
experimental magnetic dipole hyperfine coupling constants
$A_{67}(^3 \! P^o_2) = 531.996(5)$~MHz~\cite{Lurio:1962} and
$A_{67}(^3 \! P^o_1) = 609.086(2)$~MHz~\cite{Byron:1964}.
The experimental ratio $B_{63}/B_{67}$~=~1.8347(13)
of Laulainen and McDermott~\cite{LaulainenMcDermott:1969}
corresponds to $Q(^{63}{\rm Zn})=+0.275(30)$~{b}
%
%
%
(incidentally,
Laulainen and McDermott~\cite{LaulainenMcDermott:1969}
arrived at $Q(^{63}{\rm Zn})$~=~+0.29(3)~{b}).
Potential improvements could be obtained by using the measurement
of Byron \textit{et al} \cite{Byron:1964} of
$B(^3 \! P^o_1) = -18.782(8)$~MHZ
for the same $4s4p \, ^3 \! P^o_1$ state of $^{67}$Zn.
Their 65/67 ratio was $-$0.1528(3) which, combined with their
$Q(^{65}{\rm Zn})$ of 0.024(2)~{b}, corresponds
to $Q(^{67}{\rm Zn})=0.157$~{b}.
More recently the {EFGs} of Zn in solid Zn have been 
calculated in a series of papers by Haas and
collaborators~\cite{Haas:2010,Haas:2016,Haas:2017},
who employed the Density Functional Theory.
In their latest paper~\cite{Haas:2017}
using a recently developed hybrid Density Functional Theory approach,
combined with the experimental quadrupole coupling
constants measured by
Potzel~{\textit{et al}}~\cite{Potzel:1982},
and corrected for thermal effects, 
they obtained a considerably smaller value of the quadrupole moment 
$Q(^{67}{\rm Zn})$ = 0.125(5)~{b}.
{The 5~mb error limit is considered as ``may be optimistic'' in the 2017 compilation of Pyykk\"o~\cite{Pyykko:18aa}.}

In the present work magnetic hyperfine interaction constants, $A$,
and electric field gradients, $q$,
necessary for an atomic evaluation of the quadrupole moments, were calculated 
for the $ 4s 4p \,\, ^3 \! P^o_{1}$ and $ 4s 4p \,\, ^3 \! P^o_{2}$ atomic
states of the stable $^{67}$Zn isotope
using both
the non-relativistic multiconfiguration Hartree-Fock (MCHF)
method~\cite{CFFbook,FBJbook,CFFreview:2016}
and
the fully relativistic multiconfiguration Dirac-Hartree-Fock (MCDHF)
method~\cite{Grant1994,GrantBook2007,grasp2K:2013}.
{MCHF is  efficient} in capturing
electron correlation effects {while MCDHF} is necessary for correctly describing
relativistic contraction due to the mass variation, influencing the wave
function close to the nucleus.
{With this respect}, the two methods were {complementary}:
a 'DHF/HF factor' was used to correct the non-relativistic results for
the relativistic effects
{and} a 'triples correction' was used to correct the relativistic results
for the electron correlation effects arising from triple substitutions,  {a calculation that became unfeasible in the fully relativistic scheme.} 
The present work is the follow-up
of the recent measurement of the hyperfine resonances of
the $4s4p \, ^3 \! P^o_2 \rightarrow 4s5s \, ^3 \! S_1$ transition
by Wraith~\textit{et~al}~\cite{Wraith:PLB:2017},
as a {detailed} exposition {of} theoretical tools {and computational approaches}, employed to calculate magnetic fields
and electric field gradients necessary for the evaluation of nuclear
multipole moments.

The paper is divided into 
{six} sections.
Section~\ref{section.Theory}
introduces the essential elements of the multiconfiguration methods,
as well as of the theory of the hyperfine structure 
in the non-relativistic and relativistic frameworks.
Non-relativistic calculations are presented in
section~\ref{section.Non-relativistic},
%
while section~\ref{section.GRASP-JB} focuses on relativistic calculations.
In section~\ref{section.Q-4s4p} we summarize the calculations,
and we evaluate the nuclear quadrupole moment $Q(^{67}{\rm Zn})$
on the basis of eleven independent determinations of the
electric field gradients. Section~\ref{section.Conclusions} concludes the paper.

\section{Theory}
\label{section.Theory}
\subsection{Multiconfiguration methods}
In multiconfiguration methods \cite{CFFreview:2016},
the wave function, $\Psi$, for {an atomic}
state is determined as an expansion over
configuration state functions (CSFs)
\begin{equation}
\Psi = \sum_{i = 1}^{N_{\text{CSFs}}} c_i \Phi_i,
\end{equation}
where $N_{\text{CSFs}}$ is the number of CSFs in the expansion.
The CSFs are coupled anti-symmetric products of one-electron orbitals.
The expansion coefficients $c_i$ and the radial parts of the one-electron orbitals 
are determined in a self-consistent procedure by finding stationary states
of an energy functional based on a {given} Hamiltonian.
Once a radial orbital set has been determined, configuration interaction
{(CI)} 
calculations can be performed in which the expansion coefficients {only}
are determined by diagonalizing the Hamiltonian matrix. CI calculations
are simpler and faster than the self-consistent calculations and,
for this reason, the number of CSFs can be extended. 

Fully relativistic MCDHF calculations give wave functions for
fine-structure states 
and are based on the Dirac-Coulomb Hamiltonian \cite{GrantBook2007,Grant1994}.
The CSFs are obtained as $jj$-coupled and anti-symmetric products of
Dirac-orbitals.
The wave function representation in $jj$-coupling is transformed to an approximate
representation in $LSJ$-coupling,
using the methods and program developed by Gaigalas and
co-workers~\cite{Transformation,JJ2LSJ}.
The non-relativistic MCHF calculations give wave functions for $LS$ terms,
and are based on the Schr\"odinger Hamiltonian \cite{CFFbook,CFFreview:2016}.
The CSFs are obtained as $LS$-coupled and anti-symmetric products of
non-relativistic spin-orbitals.

The two methods have different strengths and weaknesses relative to the
atomic system at hand. Zinc is a fairly relativistic system for which
relativistic contraction due to the mass variation starts to get important,
especially for the calculated hyperfine constants.
These effects are captured very efficiently in the MCDHF method by the shape
of the radial orbitals. {Although the MCHF method corrected for relativistic effects through the Breit-Pauli approximation produces reliable atomic data for systems with relatively large nuclear charges~\cite{Froetal:06a}, it will never fully account for these corrections at the level of orbital optimisation~\cite{Variational}.} 
At the same time zinc is a large system with many subshells,
and electron correlation effects captured by extended CSFs expansions
are 
important for all computed properties.
Due to the restriction to $LS$ symmetry, the sizes of the CSF expansions
for MCHF calculations grow less rapidly than do the corresponding expansions
for the MCDHF calculations.  As a consequence it is possible to include
more electron correlation {excitations} in MCHF calculations.

%
%
\subsection{Hyperfine structure}
The hyperfine contribution to the Hamiltonian is represented by a multipole
expansion 
\begin{equation}
\label{H_hfs}
  H_{\rm{hfs}}=\sum_{k\geq1} {\bf T}^{(k)} \cdot {\bf M}^{(k)},
\end{equation}
where ${\bf T}^{(k)}$ and ${\bf M}^{(k)}$ are spherical tensor operators
of rank $k$ in the electronic and nuclear spaces.
The $k=1$ and $k=2$ terms represent, respectively, the magnetic dipole (M1)
and the electric quadrupole (E2) interactions. 
In non-relativistic calculations
{for an $N$-electron system} the electronic  contributions are obtained 
from the expectation values of the irreducible spherical
tensors~\cite{Hib:75b,JonWahlCFF:93} 
\begin{widetext}
\begin{eqnarray}
\label{T_1_NR}
 \mbox{\bf T} ^{(1)} = \frac{\alpha^2}{2} \sum_{j=1}^{N}
   \left\{ 2 \mbox{\bf l}^{(1)} (j) \frac{1}{r _j^3}
  -g_s \sqrt{10} [ \mbox{\bf C}^{(2)}(j) \times \mbox{\bf s}^{(1)}(j) ] ^{(1)} \frac{1}{r _j^3}
    + g_s \frac{8}{3} \pi \delta( \mbox{\bf r}_j ) \mbox{\bf s}^{(1)}(j) \right\} \; ,
\end{eqnarray}
\end{widetext}
and
\begin{equation}
\label{T_2}
 \mbox{\bf T} ^{(2)} =  - \sum_{j=1}^{N} \mathbf{C}^{(2)}(j) \frac{1}{r_j^{3}} \; .
\end{equation}
In the fully relativistic approach,
the magnetic dipole electronic tensor reduces to a single
term~\cite{LinRos:74a,Jonetal:96c} 
\begin{equation}
\label{T1_R}
 \mbox{\bf T} ^{(1)} = -i \alpha \sum_{j=1}^{N} 
 \left( {\bm{ \alpha }}_j \cdot
  \mbox{\bf l}_j \;  \mbox{\bf C}^{(1)}(j)  \right)  \frac{1}{ r_j^2} \; .
\end{equation}
The electronic contribution for the magnetic dipole interaction is combined with the nuclear spin $I = 5/2$ and the measured nuclear magnetic dipole
moment $\mu = 0.875479 \, \mu_N$ \cite{NJStone:2016} to give the magnetic dipole hyperfine interaction constant, $A$, for the $4s4p \, ^3 \! P^o_{1,2}$ states in $^{67}$Zn. 
{The electric field gradient (EFG), also denoted $q$~\cite{Pyykko:97c}, is obtained from the reduced matrix element of the operator~(\ref{T_2}) using the electronic wave function of the considered electronic state (see \cite{Jonetal:96c,Bieron:e-N:2015} for details). It corresponds to the electronic part of the electric quadrupole hyperfine interaction constant, $B$. The latter, expressed in MHz, can be calculated  using the following equation
\begin{equation}
\label{B_frequency_barns_3}
B/\mbox{MHz } =  234.9646 \, (q/a_0^{-3}) (Q/\mbox{b}) \; ,
\end{equation}
where the EFG ($q$) and the nuclear quadrupole moment ($Q$) are expressed in $a_0^{-3}$ and barns, respectively.
 Instead of reporting  $q$, we will monitor in the present work the related $B/Q \propto q$ ratio values (in MHz/b). }

\section{Non-relativistic calculations}
\label{section.Non-relativistic}

\subsection{MCHF calculations}
\label{section.MCHF.4s4p}
The MCHF calculations were performed using the Atomic Structure Package  
(ATSP2K) \cite{ATSP2k:2007}.
As a starting point a Hartree-Fock (HF) calculation was performed for
$ 4s 4p \,\, ^3 \! P^o$. The HF calculation was followed by a sequence of calculations describing valence-valence and core-valence electron correlation effects. The CSF expansions for these calculations were obtained by allowing 
single {(S)} 
and double {(D)}
substitutions from $ 2s^22p^63s^23p^63d^{10}4s 4p \,\, ^3 \! P^o$ to increasing active sets of orbitals with the restriction that 
there is at most one substitution from the core shells. {The $1s$-shell} is kept closed in all calculations. These expansions are referred to as all singles and restricted doubles (SrD) expansions. 
The active sets are denoted by giving the highest orbital of each symmetry. For example, $\{5s5p4d4f\}$ denotes the orbital set that includes 
the orbitals $1s,2s,3s,4s,5s,2p,3p,4p,5p,3d,4d,4f$. 
In the MCHF calculations the HF orbitals were kept frozen and the remaining orbitals were optimized together.
The MCHF calculations were followed by a CI calculation based on the largest orbital set. The CI calculation describes valence-valence {(VV)}, core-valence {(CV)} and core-core {(CC)} correlation effects and includes CSFs obtained by all single and double (SD) substitutions. Whereas SD expansions describe the major corrections to the wave function, it is known that CSFs obtained from triple (T) substitutions are important for hyperfine structures {\cite{PorDer:2006a}}. The effects of the T substitutions were accounted for in CI,  by augmenting the largest SD expansion with expansions obtained by T substitutions to increasing orbital sets. All calculations are summarized in \tablename{~\ref{table.summaryMCHF}}.

To correct for the relativistic contraction due to the mass variation, Dirac-Hartree-Fock (DHF) calculations were also performed and the final SD+T values were multiplied with the DHF/HF ratio. This correction will be discussed in more detail in the next section.
From \tablename{~\ref{table.summaryMCHF}}
we see that valence-valence and core-valence effects, as described by SrD expansions, increase the absolute values of all computed hyperfine parameters. The increase is around 30~\% for the $A$ constant and 60~\% for the {electric} field gradient {$q \propto B/Q $}. The changes are well converged and are consistent {with a contraction of the wave function when accounting for core-valence correlation as observed in \cite{Linetal:76a,Sal:84a}}. 
The effect of unrestricted D substitutions, resulting in CSFs describing also core-core correlation, is to decrease the absolute values of the computed hyperfine parameters. The CSFs resulting from the unrestricted
{double}
substitutions can be shown to have small effects on the hyperfine parameters by themselves. Instead the effects are indirect, changing or effectively diluting the mixing coefficients of the more important CSFs describing core-valence effects~\cite{Engels1993,Engels1996,PJMRG,MRGPJ}.

\begin{table*}[!htbp]
\caption{MCHF calculations of $A$ (MHz) and {~$B/Q$ (MHz/b)~}of $ 4s 4p \,\, ^3 \! P^o_{1,2}$ in $^{67}$Zn 
($I^{\pi}=5/2^-$ and $\mu_{\text{expt}}=0.875479(9)\,\mu_N$).
SrD denote all single and restricted double expansions to increasing active orbitals sets. SD denote single and double expansions to the largest orbital set. 
{Single and restricted Double (SrD) expansions, with at most one substitution from the core shells, allow the inclusion of valence (VV) and core-valence (CV) effects (see text). {Core-core correlation} (CC) is included through unrestricted D substitutions.} 
SD+T
denote expansions where the largest SD expansion has been augmented by expansions from T substitutions to increasing active orbital sets.
{The largest SD+T results were scaled by the DHF/HF ratio factor in the line labeled MCHF$\times$DHF/HF.}
HF = uncorrelated Hartree-Fock values;
DHF = uncorrelated Dirac-Hartree-Fock values.
{$N_{\text{CSFs}}$} is the number of CSFs in the expansion.}
\label{table.summaryMCHF}
\begin{tabular}{lrdddd}
\hline\noalign{\smallskip}
\hline\noalign{\smallskip}
    &  & \multicolumn{2}{c}{$^3 \! P^o_{1}$} & \multicolumn{2}{c}{$^3 \! P^o_{2}$} \\
 Label             & 
 \multicolumn{1}{c}{ $N_{\text{CSFs}}$} & 
 \multicolumn{1}{r}{         $A$ (MHz) } & 
 \multicolumn{1}{r}{                    {~$B/Q$ (MHz/b)~}}& 
 \multicolumn{1}{r}{                                  $A$ (MHz)} &
 \multicolumn{1}{r}{                                              {~$B/Q$ (MHz/b)~}}\\
\colrule
HF                 &       & 412.72451  &   -93.033 &  373.00296   & 186.066 \\
DHF                 &       & 473.40239  &  -100.373 & 419.93437   & 192.924 \\
\hline\noalign{\smallskip}
SrD (VV+CV) \\         
$5s5p4d4f$           &   404 & 471.978 & -120.16 & 429.223 & 240.33 \\
$6s6p5d5f5g$         &  1593 & 507.266 & -136.52 & 459.667 & 273.04 \\
$7s7p6d6f6g6h$       &  3872 & 526.518 & -142.07 & 476.132 & 284.15 \\
$8s8p7d7f7g7h$       &  7232 & 536.870 & -146.85 & 484.970 & 293.70 \\
$9s9p8d8f8g8h $      & 11673 & 541.017 & -148.95 & 488.430 & 297.90 \\
$10s10p9d9f9g9h$     & 17195 & 542.624 & -148.27 & 489.789 & 296.55 \\
$11s11p10d10f10g10h$ & 23798 & 542.926 & -148.44 & 490.022 & 296.89 \\
\hline\noalign{\smallskip}
SD (VV+CV+CC) \\   
 $11s11p10d10f10g10h$ & 44546 & 521.477  & -137.738  & 470.799  & 275.476  \\

\hline
SD+T (VV+CV+CC) &    & & & & \\  
$5s5p4d4f$           & 92810   & 533.150 & -141.59 & 481.192 & 283.18 \\
$6s6p5d5f5g$         & 225457  & 540.485 & -144.86 & 487.481 & 289.73\\ 
$7s7p6d6f6g$         & 446457  & 544.960 & -147.53 & 491.050 & 295.06 \\ 
$8s8p7d7f7g$         & 761267  & 551.678 & -151.68 & 496.389 & 303.36\\ 
$9s9p8d8f8g$         & 1175344 & 553.437 & -152.988 & 497.691 & 305.976 \\
\hline
 MCHF $\times$ DHF/HF &  &       634.802  &   -165.058 & 560.310    & 317.254 \\
Expt.                &      & 609.086^\text{a}      &          & 531.987^\text{b}   \\
\hline
\hline
\end{tabular}
\begin{flushleft}
\small{$^\text{a}$Byron~\textit{et~al}~\cite{Byron:1964}.} \\
\small{$^\text{b}$Lurio~\cite{Lurio:1962}.}
\end{flushleft}
\end{table*}
%
Finally, the effect of the T substitutions is to increase the absolute
values of the hyperfine constants. Again, the effect is mainly indirect,
affecting the expansion coefficients of the important singly
excited CSFs:
$2s^22p^63snl3p^63d^{10}4s4p \, ^3 \! P^o$ and \\
$2snl2p^63s^23p^63d^{10}4s4p \, ^3 \! P^o$, describing spin-polarization; \\
$2s^22p^63s^23p^5nl3d^{10}4s4p \, ^3 \! P^o$ and \\
$2s^22p^5nl3s^23p^63d^{10}4s4p \, ^3 \! P^o$, the last two
describing orbital-polarization~\cite{Engels:1987,MRGPJ}.
The latter effects will be analyzed in more detail in
{the section 
\ref{subsection.PCFI}
below.}
The general convergence trends and behavior with respect to different correlation effects are consistent with the ones {found for} other similar systems \cite{PJMRG}. It is interesting to note that the effects discussed above are partly canceling. Thus it is better to include only the valence-valence and core-valence effects than the  valence-valence, core-valence and core-core effects. If the core-core effects are included, then also the effects of the T substitutions should be accounted for. The final $A$ constants for the $ 4s 4p \,\, ^3 \! P^o_{1,2}$ states differ from the experimental values by 1.8 \% and \mbox{5.2 \%}, respectively.
\subsection{DHF/HF correction}
\label{section.DHF-HF-correction}

\tablename{~\ref{table.orbital.radii}} presents expectation values $ \langle r \rangle_{nl} $  and 
$ \langle r \rangle_{n \kappa} $ of spectroscopic orbitals obtained in zeroth-order (no electron correlation) non-relativistic Hartree-Fock (HF) and relativistic Dirac-Hartree-Fock (DHF) approximations, 
{where $\kappa = -(l+1)$ for $j=l+1/2$ and  $\kappa = +l$ for $j=l-1/2$.} For all spectroscopic orbitals but $3d$, the direct relativistic contraction due to the mass variation dominates the indirect one, induced by the relativistic charge redistribution~\cite{Des:2002a}. The differences in radii of correlation orbitals are more complex. They usually
reflect specific correlation effects, targeted in the self-consistent-field optimisation strategies~\cite{CFFreview:2016}. For the M1 hyperfine interaction, a detailed comparison of  the non-relativistic expectation values of eq. (\ref{T_1_NR}) and of the relativistic ones of eq. (\ref{T1_R})  in terms of single-electron orbitals contributions is not easy. The global effect in the single configuration approximation
is to produce large DHF/HF ratios of M1 and E2 hyperfine constants, as illustrated by the first two lines of \tablename{~\ref{table.summaryMCHF}}.
As can be seen, the relativistic effect is much larger for the $A$ constants than for the {EFG} values, which can be explained by the {contact interaction that appears  in the non-relativistic expression for the M1 interaction with the three-dimensional delta function (see {the}
last term of eq.~(\ref{T_1_NR})). Although the corresponding relativistic expression of eq.~(\ref{T1_R}) does not contain a contact operator, the tensorial structure of {the}
relativistic operator indicates that it is highly biased
{towards}
the behavior of the wave function close to the nucleus where the relativistic contraction effects are the most important.}
%
\begin{table*}[!htbp]
%
\caption{HF $ \langle r \rangle_{nl} $ vs
         DHF $ \langle r \rangle_{n \kappa} $ orbital radii ($a_0$).
HF calculation for the  $ 4s 4p \,\, ^3 \! P^o$ term.
DHF calculation optimised on the
$ 4s 4p \,\, ^3 \! P^o_{0,1,2}$ states together.
Notation:
'$nl$' = HF orbital;
'$nl+$' = DHF orbital with negative $\kappa$
($1s$, $2s$, $3s$, $4s$, $2p$, $3p$, $3d$, $4p$);
'nl$-$' = DHF orbital with positive $\kappa$ 
($2p-$, $3p-$, $3d-$, $4p-$). The table illustrate{s}
the direct relativistic contraction of $s$ orbitals 
due to mass variation.}
\label{table.orbital.radii}  
\begin{tabular}{lcccccccc}
%
%
%
%
\colrule 
    & $1s$      & $2s$      & $2p$      & $3s$      & $3p$      & $3d$      & $4s$      & $4p$      \\
$nl$  & 0.05108 & 0.22878 & 0.19951 & 0.69107 & 0.71948 & 0.87132 & 2.77730 & 3.80035 \\
$nl+$ & 0.05028 & 0.22498 & 0.19912 & 0.68097 & 0.71824 & 0.87909 & 2.72995 & 3.79473 \\
$nl-$ &       &         & 0.19578 &         & 0.70820 & 0.87169 &         & 3.76246 \\
%
%
%
\colrule 
\end{tabular}
\end{table*}
{Relativistic effects in nuclear quadrupole couplings have been investigated by Pyykk\"o and Seth \cite{Pyykko:97c} who estimated relativistic correction factors $C$ for {EFGs} due to valence $p$ electrons from one-electron matrix elements $q_{+ +}$, $q_{+ -}$ and {$q_{- -}$}. The $q_{- -}$ combination has $j=1/2$ for $l=1$ and corresponds to a spherical charge distribution. It will therefore not contribute to the {EFG}, oppositely to $q_{+ +}$ and $q_{+ -}$.} 
{These $C$ factors can be used to scale the non-relativistic {EFG} values.}
{They} {have been estimated in the quasirelativistic (QR) approximation (no fine-structure splitting), using hydrogen-like (H) and Dirac-Fock (DF) expectation values, or the Casimir's $n$-independent formulae (Cas).}
{They} {are reported in the first three lines of~\tablename{~\ref{table.DHF-DF-factor}}. Beyond the QR approximation, correction factors can be estimated for the fine-structure levels of light atoms 
 by taking the right combination of the $C$ coefficients, 
 $\left( -1/3 \, C_{++} + 4/3 \, C_{+-} \right) $ and $C_{++}$ for  $J=1$ and $J=2$, respectively, with  $C_{++} = 1.02556$ and $C_{+-} = 1.06177$~\cite{Pyykko:97c}.} These factors are also reported in {the~\tablename{~\ref{table.DHF-DF-factor}}},
 and compared with the DHF/HF  ratios estimated from the {EFG} values reported 
 in~\tablename{~\ref{table.summaryMCHF}}.

These ratios,  {obtained} in the single configuration picture,
 may be used to scale the multiconfiguration results,
 {as it is done {in} the line MCHF$\times$DHF/HF of \tablename{~\ref{table.summaryMCHF}}},
{with the underlying assumption that cross terms
between relativistic contraction and electron correlation are negligible.}
Looking at the differences in ratios between different methods, we infer that application of the DHF/HF corrective ratio induces an uncertainty  of at most 2-3~\% for the  electric field gradient $q$ of the $J=1$ state. The uncertainty is smaller for the $J=2$ state.
       
%

%


\begin{table*}[!htbp]
%
\caption{{Relativistic correction factors for {EFG} estimated with different methods. Quasirelativistic correction factors reported are taken from Ref.~\cite{Pyykko:97c} using Dirac-Fock (DF), Hydrogen-like (H) expectation values or the Casimir's $n$-independent formulae (Cas) (see Ref.~\cite{Pyykko:97c} for more details.) The $J$-dependent correction factors are either calculated following the procedure  {outlined}
in the  conclusion section of Ref.~\cite{Pyykko:97c}, or from the DHF/HF ratio of {EFG} values (this work).} 
"\text{\bf{+}}"~=~$p_{3/2}$ orbital; 
"\text{\bf{--}}"~=~$p_{1/2}$ orbital.}
\label{table.DHF-DF-factor}  
\begin{tabular}{lll}
%
%
\colrule 

QR approach~~~~ &1.05468  & \phantom{m} {$C_{\rm{QR}}$(Cas)}~: Casimir's n-independent formula~{\cite{Pyykko:97c}}\\
 &1.05776  & \phantom{m} {$C_{\rm{QR}}$(H)}~: H-like~{\cite{Pyykko:97c}}\\
 &1.04970  & \phantom{m} {$C_{\rm{QR}}$(DF)}~: DHF~{\cite{Pyykko:97c}} \\
 &  \\
 $J=1$ &1.07384  & \phantom{m} {calculated from~\cite{Pyykko:97c}  $(-1/3 \, C_{++} + 4/3 \, C_{+-})$} \\
       &1.07890  & \phantom{m} {DHF/HF ratio}, this work {(1st line of \tablename{~\ref{table.summaryMCHF}}).} \\
       & & \\
 $J=2$ &1.02556  & \phantom{m} {calculated from~\cite{Pyykko:97c}  $(C_{++})$} \\
       &1.03686  & \phantom{m} {DHF/HF ratio}, this work {(2nd line of \tablename{~\ref{table.summaryMCHF}}).} \\
%
\colrule 
\end{tabular}
\end{table*}
%


\subsection{Contributions to $A$ and {EFG} from different classes of orbital
 substitutions}
\label{subsection.PCFI}

\begin{table}[!htbp]
\caption{The effect on $A$ and {$B/Q$~($\propto$~EFG)} of $ 4s 4p \,\, ^3 \! P^o_{1,2}$
         from different classes of orbital substitutions.
	 Analysis for the final SD calculation.
	 See text for details of the notation and for a discussion about the importance of different classes.}
\label{table.summaryPCFI}  
\begin{tabular}{ldddd}
\hline\noalign{\smallskip}
\hline\noalign{\smallskip}
    &  \multicolumn{2}{c}{$^3 \! P^o_{1}$} & \multicolumn{2}{c}{$^3 \! P^o_{2}$} \\
 Label             & 
 \multicolumn{1}{r}{         $A$ (MHz) } & 
 \multicolumn{1}{r}{                     {~$B/Q$ (MHz/b)~}}& 
 \multicolumn{1}{r}{                                  $A$ (MHz)} &
 \multicolumn{1}{r}{                                              {~$B/Q$ (MHz/b)~}}\\
\colrule
HF                       & 412.72  &   -93.033 & 373.00 & 186.06 \\
\hline\noalign{\smallskip}
$vv$   & 411.07  &  -92.28  &  371.67  &  184.56  \\
$3dv$  & 479.96  & -109.23  &  433.48  &  218.46  \\
$3d$   & 479.93  & -112.88  &  434.33  &  225.76  \\
$3d3d$ & 459.77  & -106.35  &  416.60  &  212.71  \\
$3pv$  & 470.26  & -109.49  &  425.92  &  218.98  \\
$3p3d$ & 469.76  & -109.55  &  425.52  &  219.10  \\
$3p$   & 477.31  & -133.89  &  428.89  &  267.78  \\
$3p3p$ & 475.54  & -133.21  &  427.30  &  266.43  \\
$3sv$  & 478.39  & -133.58  &  429.93  &  267.16  \\
$3s3d$ & 479.19  & -133.96  &  430.54  &  267.92  \\
$3s3p$ & 479.35  & -134.17  &  430.60  &  268.34  \\
$3s$   & 506.62  & -134.16  &  457.86  &  268.32  \\
$3s3s$ & 506.11  & -134.07  &  457.38  &  268.15  \\
$2pv$  & 509.65  & -135.20  &  460.48  &  270.40  \\
$2p3d$ & 509.88  & -135.52  &  460.61  &  271.05  \\
$2p3p$ & 509.72  & -135.34  &  460.47  &  270.68  \\
$2p3s$ & 509.71  & -135.44  &  460.41  &  270.89  \\
$2p$   & 512.05  & -139.21  &  461.24  &  278.43  \\
$2p2p$ & 509.03  & -137.95  &  458.45  &  275.90  \\
$2sv$  & 510.21  & -138.12  &  459.55  &  276.25  \\
$2s3d$ & 510.53  & -138.08  &  459.86  &  276.16  \\
$2s3p$ & 510.91  & -138.08  &  460.24  &  276.16  \\
$2s3s$ & 511.00  & -138.06  &  460.33  &  276.13  \\
$2s2p$ & 511.36  & -138.15  &  460.66  &  276.30  \\
$2s$   & 521.91  & -137.89  &  471.20  &  275.78  \\
$2s2s$ & 521.47  & -137.73  &  470.79  &  275.47  \\
\hline
\hline
\end{tabular}                                                      
\end{table}                                                         


The uncertainties of the computed $A$ constants and electric field gradients $q$ are to a large extent determined by the size of the cancellation effects~{\cite{CarGod:2011a,CarGod:2011b}}.
In the non-relativistic formalism the $A$ constants are computed based on the operator in \mbox{eq. (3)} and are the sums of three terms $A_l$, $A_{sd}$, $A_c$, orbital, spin-dipolar, and Fermi contact term, {respectively}.
At the HF level we have (in MHz):
$A_l = 33.06$, $A_{sd} = 33.10$, $A_c = 346.56$, and
$A_l = \, 33.06$, $A_{sd} = -6.62$, $A_c = 346.56$, for
$4s4p \, ^3 \! P^o_{1}$ and $4s4p \, ^3 \! P^o_{2}$, respectively.
It is seen that the Fermi contact term dominates, but 
this contribution is partly canceled by the spin-dipolar contribution
for the $J=2$ state. 
Based on this simple observation we may expect that the computed $A$ constant is less accurate for {the $J=2$ state.} 
To shed light on the sensitivity of $A$ and ${B/Q}$ to electron
correlation effects we analyze the contributions to these parameters
from different classes of orbital substitutions.
Given the $\{11s11p10d10f10g10h\}$ orbital set, the  $A$ constants and {$B/Q$ ratio values}
are computed from 
accumulated CSF expansions that result from allowing  {single} 
and  {double} substitutions
from deeper and deeper lying orbitals of the $2s^22p^63s^23p^63d^{10}4s4p$
reference configuration. The results are presented in 
\tablename{~\ref{table.summaryPCFI}}.
The accumulated CSF expansions are denoted by the innermost orbitals from
which the substitutions are allowed.
For example, $3d3d$ denotes the accumulated CSF expansion  
that is obtained by allowing the substitutions
\[
vv \to nln'l',~~3dv \to nln'l',~~3d \to nl,~~3d3d \to nln'l'\phantom{,~~3pv \to nln'l'}
\] 
whereas $3pv$ denotes the accumulated CSF expansion obtained from the
substitutions
%
%
\[
vv \to nln'l' 
\] 
\vspace{-28pt}
\[
3dv \to nln'l'
\] 
\vspace{-28pt}
\[
3d \to nl
\] 
\vspace{-28pt}
\[
3d3d \to nln'l'
\] 
\vspace{-28pt}
\[
3pv \to nln'l'
\] 
%
where $nl,\,n'l' \in \{11s11p10d10f10g10h\}$.
By comparing the results for $3d3d$ and $3pv$ we can infer how large are
the contributions from CSFs obtained from the
$3pv \to  nln'l'$ substitutions.
From \tablename{~\ref{table.summaryPCFI}}
one can see that CSFs obtained from $3dv \to  nln'l'$ substitutions
describing core-valence correlation are very important for both $A$ and {$B/Q$}.
One can also see that CSFs obtained from $3s, \, 2s \to  nl$ substitutions describing 
spin-polarization are important for the $A$ parameters whereas CSFs obtained
from $3d, \, 3p, \, 2p \to  nl$ substitutions describing orbital-polarization are important
for the $q$ parameters. One further notes that the effects of CSFs
from {single} substitutions are often canceled by {those} of CSFs from {double} substitutions.
Of particular importance are the effects from $3d3d \to  nln'l'$. The corresponding CSFs do not directly contribute to the
hyperfine parameters but they are important for the total wave function,
lowering, or diluting, the effects of the other CSFs (compare the discussion in 
the previous section).
{The accuracy of the calculated $A$ {constant and $(B/Q)$ ratio values}  is to a large extent determined by the fact that they}
{result}
from {a} {summation} of a number of
canceling contributions.
{We refer to chapter~8 of~\cite{FBJbook}} for a general discussion about spin- and orbital-polarization effects.

\section{Multiconfiguration Dirac-Hartree-Fock/RCI calculations}
\label{section.GRASP-JB}

Two different approaches were used for the
$ 4s 4p \,\, ^3 \! P^o_{1,2}$ states.
In the first approach, called OL1 (Optimal Level~1)
the wave functions for the
$ 4s 4p \,\, ^3 \! P^o_{2}$ state were optimised for {a single} state, i.e.~the
$ 4s 4p \,\, ^3 \! P^o_{2}$ {level} itself.
In the second approach (called OL4) the wave functions were generated
with the Extended Optimal Level~\cite{grasp89} form
of the variational functional,
built from all 4 states of the $4s4p$ configuration
($ 4s 4p \,\,\, ^3 \! P^o_{0}, \, ^3 \! P^o_{1}, \, ^3 \! P^o_{2}, \, ^1\! P^o_{1}$).
The full description of numerical methods, virtual orbital sets,
electron substitutions, and other details of the computations,
can be found
in~\cite{grasp2K:2013,CFFreview:2016,%
BieronTiQ1999,BieronBeF1999,BieronAu2009,JonssonBieron2010,Bieron:e-N:2015}.

\subsection{Optimal Level calculations for the
	$4s4p \, ^{3} \! P^{o}_{2}$ state}
\label{subsection.GRASP-4s4pOL1}

As mentioned {above, the first approach (OL1) targets the optimisation of the single 
state $ 4s 4p \,\, ^3 \! P^o_{2}$ wave function.}
%
The spectroscopic orbitals $1s2sp3spd4sp$ were generated in 
Dirac-Hartree-Fock (DHF) mode, i.e.~without correlation (virtual) orbitals,
and were frozen through all further steps.
Five layers of virtual orbitals~\cite{BieronAu2009}
of $s$,$p$,$d$,$f$,$g$,$h$ angular symmetries
were  {sequentially}
generated {by including single and double substitutions (SD) for}
the first two layers and {single and restricted double substitutions (SrD) for the}
{third, fourth, and fifth layer.}

The occupied shells were {successively}
opened for substitutions into virtual set,
starting with $4sp$, followed by $3spd$, and then by $2sp$.
The $1s$ shell was kept closed {in all calculations}.
The {multiconfiguration self-consistent-field optimisation step}
 was followed by configuration interaction~(RCI) calculations, 
 in which {CSF} expansions were appended
with configurations arising from subsets of
unrestricted single and double (SD) substitutions,
or with (subsets of) unrestricted single, double, and triple (SDT)
substitutions.

%
%
%
\begin{figure}
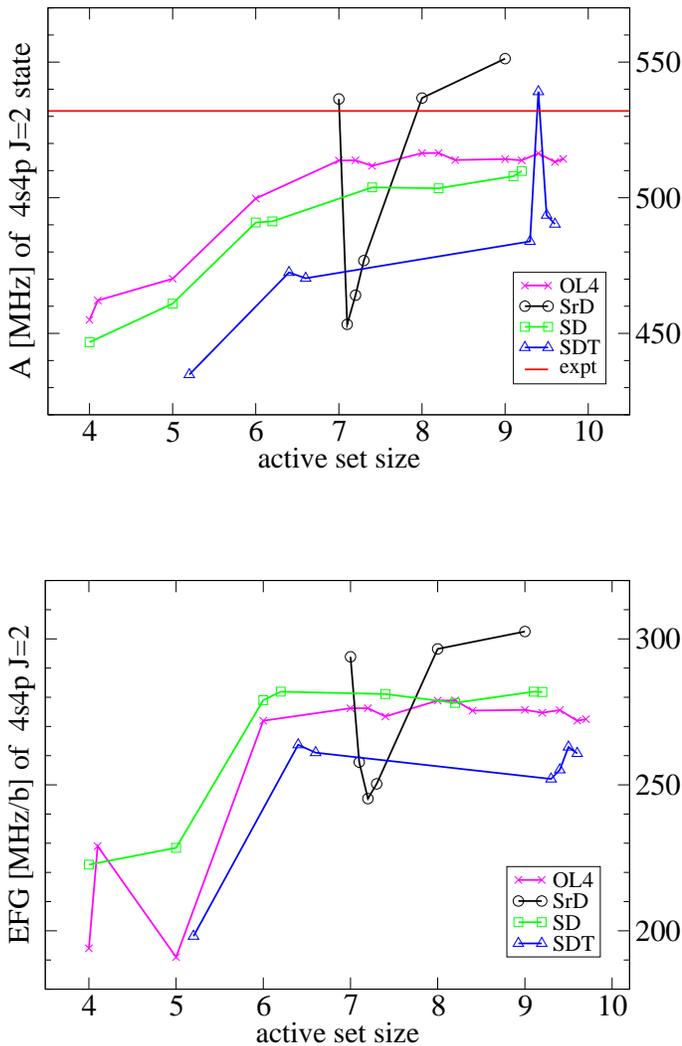

\includegraphics[width=.50\textwidth]{4s4pj2-ol1-a.eps}
\vspace{28pt} \phantom{ppp}
\hspace{-44pt} \phantom{ppp}
\includegraphics[width=.50\textwidth]{4s4pj2-ol1-efg.eps}
\caption{(Color online) Hyperfine constant $A( 4s 4p \,\, ^3 \! P^o_{2})$~(MHz)
(curves in the upper graph of the figure)
and $B/Q$ ratio~(MHz/b) of the $ 4s 4p \,\, ^3 \! P^o_{2} $ state
(curves in the lower graph of the figure), obtained in several approximations.
{Each integer value on the abscissa axis represents
the maximal principal quantum number of the virtual orbital set
for a particular multiconfiguration expansion.
The fractional values represent approximations, 
where multiconfiguration expansions were appended
with subsets of SD or SDT expansions.
The straight horizontal line (red online) represents the experimental value
$ A ( ^3 \! P^o_{2} )  = 531.987(5) $~MHz.
More details are provided in text.}}
\label{figure.4s4pj2-ol1}
\end{figure}
%


Figure~\ref{figure.4s4pj2-ol1} shows the dependence of the
magnetic dipole hyperfine constant $ A $~(MHz),
(curves in the upper {graph} of the figure),
and
{the $B/Q$ ratio~(MHz/b), proportional to EFG, of}
the $ 4s 4p \,\, ^3 \! P^o_{2} $ state
(curves in the lower  {graph} of the figure),
on the size of the multiconfiguration expansion.
All lines in both graphs are drawn only for the guidance of the eyes.
The results of the calculations are represented by  {several}
symbols
described in the following paragraph.
Each {integer} value on the abscissa axis represents
the maximal principal quantum number of the virtual orbital set
for a particular multiconfiguration expansion.
The fractional values represent approximations,
where {CSF} expansions were appended
with subsets of SD or SDT expansions.
{In these configuration interaction calculations}, these subsets
were generated in the following ways: 
the occupied orbitals were systematically opened for SD and SDT substitutions;
{the size of the virtual orbital set was systematically increased
 for SD substitutions, until the expectation values saturated with respect to
 the size of the virtual orbital set;
 then the size of the virtual orbital set was systematically increased for SDT
 substitutions.
{The convergence of the SDT results was not reached {since}
larger SDT multiconfiguration expansions would exceed the capacity of the
computer systems at our disposal (6x96~CPU~@~2.4GHz with 6x256~GB~RAM).}}

A stepwise, systematic increase of different classes of substitutions makes it
possible to identify those classes which bring about considerable contributions
to the expectation value(s), as well as to 
{quantify} these contributions. Those with sizeable contributions were later included
in the final configuration interaction calculations.
Four curves in Figure~\ref{figure.4s4pj2-ol1} represent the following {correlation models}: 
\begin{itemize}
\item 
\vspace{-6pt}
circles (black online) = single and restricted double substitutions (SrD); 
\vspace{-6pt}
\item
squares (green online) = unrestricted single and double (SD) substitutions; 
\vspace{-6pt}
\item
triangles (blue online) = single and double and triple (SDT) substitutions; 
\vspace{-6pt}
\item 
crosses (magenta online) = results of the OL4 calculation, described in 
     section~\ref{subsection.GRASP-4s4pOL4} below. 
\vspace{-6pt}
\end{itemize}
The curves with circles (black online) represent the initial phase of
the calculations, where the third, fourth, and fifth {layers} of virtual orbitals were generated with
single and restricted double substitutions (SrD).
These results were not corrected for unrestricted double (SD), nor for triple (SDT) substitutions.
The curves with triangles (blue online) represent the configuration interaction
calculations, where unrestricted double and triple (SDT) substitutions were
included. However, due to the limitations of available computer resources
the triple substitutions were limited to substitutions from
$4s$,$4p$ occupied orbitals
to one layer of virtual orbitals,
or substitutions from $3s$,$3p$,$3d$,$4s$,$4p$ occupied orbitals 
to two layers of virtual orbitals.
The {oscillations} of the blue curves is a clear evidence that 
the triple substitutions were not saturated in these calculations.

%
%
The SD and OL4 curves,
{with squares (green online) and crosses (magenta online)} in
both upper and lower graphs, respectively,
in Figure~\ref{figure.4s4pj2-ol1} 
represent the values corrected for the triple substitutions
in a systematic manner:  the triple substitutions
were accounted for
with an additive correction computed with the 
non-relativistic
Hartree-Fock
program ATSP2K~\cite{ATSP2k:2007}
(see section~\ref{subsection.triples-corrections} below).

The straight horizontal line (red online) 
accross the upper part of the upper graph
in Figure~\ref{figure.4s4pj2-ol1} represents the {experimental} 
magnetic dipole hyperfine constant $A = 531.987(5)$~MHz
for the $ 4s 4p \,\, ^3 \! P^o_{2} $
state of the $^{67}$Zn~isotope~\cite{Lurio:1962}.

The end {products} of the calculations described in the present section
{are} the magnetic dipole hyperfine  constant, {$A = 509.861$~MHz}
and the 
{$B/Q=281.799$~MHz/b} ratio represented {in
Figure~\ref{figure.4s4pj2-ol1}} by the {points} 
at the right hand side {ends} of the
curve with squares (green online) on the upper {and lower graphs, respectively.}
These values were obtained from
the configuration interaction calculation with
single and double substitutions from $2sp3spd4sp$ occupied orbitals
to five layers of virtual orbitals
(the largest size of the virtual orbital set generated in this approximation).
These results, corrected for triple substitutions {as} described in the
section~\ref{subsection.triples-corrections} below,
were considered final in the single-reference calculations 
described in the present section. {They} are quoted in \tablename{~\ref{table.summaryAQZn}}
in line marked 'MCDHF-SD-SR-OL1+t(MCHF)'.

The scatter of points at the right hand side end of the four curves 
presented in the 
Figure~\ref{figure.4s4pj2-ol1} 
and the oscillations of the individual curves
{could} serve as {a guideline for estimating the error bars of the theoretical
{EFG} contribution to $B$, and indirectly to {$Q \propto B/q$.}
In the present paper however, the error {bars have} been estimated with
a somewhat more reliable procedure described in the
section~\ref{section.Q-4s4p} below.

\subsection{Additive corrections for triple substitutions}
\label{subsection.triples-corrections}

\begin{table*}[!htbp]
%
\caption{Corrections for triple substitutions, (SDT$-$SD),
calculated  
for hyperfine constant
$A$ (MHz)
and {~$B/Q$~(MHz/b)}
in $ 4s 4p \,\, ^3 \! P^o_{1,2}$ states.
}
\label{table.triples-correctionsAEFG}
\begin{tabular}{ldddd}
%
%
%
%
\colrule 
%
%
              & \multicolumn{1}{l}{ \phantom{m} $A(J=1)$} 
              & \multicolumn{1}{l}{             {$B/Q$}$(J=1)$}       
              & \multicolumn{1}{l}{ \phantom{m} $A(J=2)$}       
              & \multicolumn{1}{l}{ \phantom{mm} {$B/Q$}$(J=2)$} \\     
\colrule 
SD    & 521.4771     & $-$137.7380     &  470.7987    & 275.4760     \\
SD+T  & 553.4370     & $-$152.9882     &  497.6906    & 305.9764     \\
SDT$-$SD & 31.96     & $-$ 15.25\phantom{82} & 26.89  &  30.50     \\
%
\colrule 
\end{tabular}
\end{table*}
%


If the contribution of triple substitutions is small, it may be approximately
assumed as an additive correction, approximately independent of relativity,
and may be computed in the non-relativistic framework as the difference
between the values obtained with and without triple substitutions, respectively.
As an example,
the correction ($31.96$~MHz) for the magnetic dipole hyperfine constant
$A(4s 4p \,\, ^3 \! P^o_{1})$
in \tablename{~\ref{table.triples-correctionsAEFG}}
was evaluated as the difference between
the value calculated in the SD+T approximation
($A = 553.4370   $~MHz)
and
the value calculated in the SD approximation
($A = 521.4771   $~MHz).
Analogous differences were assumed as
triple contributions for the $A(4s 4p \,\, ^3 \! P^o_{2})$ constant,
as well as for the {$B/Q$} ratio-values for both states.

\subsection{Extended Optimal Level calculations for the
        $4s4p \, ^{3} \! P^{o}_{1}$ and $4s4p \, ^{3} \! P^{o}_{2}$ states}
\label{subsection.GRASP-4s4pOL4}

The calculations {described} in this section were performed in a similar manner as those
{presented} in section~\ref{subsection.GRASP-4s4pOL1},
with one significant {difference}:
wave functions were optimised for all {four} states of $4s4p$ configuration
($ 4s 4p \,\,\, ^3 \! P^o_{0}, \, ^3 \! P^o_{1}, \, ^3 \! P^o_{2}, \, ^1\! P^o_{1}$)
in the Extended Optimal Level (OL4) approach~\cite{grasp89},
with equal weights.
The calculations of hyperfine $A$ and {EFG} factors
for $ 4s 4p \,\, ^3 \! P^o_{1,2}$ states presented
in this section are computationally more demanding
than those for $ 4s 4p \,\, ^3 \! P^o_{2}$ state presented
in section~\ref{subsection.GRASP-4s4pOL1}.
The $4s4p$ configuration splits into four levels
($^3 \! P^o_0$, $^3 \! P^o_1$, $^3 \! P^o_2$, $^1P^o_1$)
and there are two levels of $J=1$ symmetry.
The singlet {$^1P_1^o$} state interacts considerably with the triplet $^3 \! P^o_1$ state,
and in such situations optimisation on all close lying levels 
often yields a better balance of {states involved in} configuration mixings.
However, the multiconfiguration expansions are larger, and self-consistent-field
process requires considerably more computer resources.
The end product{s} of the calculations described in the present section
are the hyperfine $A$ {constants and $B/Q$ ratios}
for the $ 4s 4p \,\, ^3 \! P^o_{1,2}$ states,
obtained from
the configuration interaction calculation with
single and double substitutions from $2sp3spd4sp$ occupied orbitals
to five layers of virtual orbitals
(the largest size of the virtual orbital set generated in this approximation).
%
These results, corrected for triple substitutions {as} described in the
section~\ref{subsection.triples-corrections} above,
were considered final in the Extended Optimal Level calculations 
and they are quoted in \tablename{~\ref{table.summaryAQZn}}
in lines marked `MCDHF-SD-SR-OL4+t(MCHF)'
(separately for 
$ 4s 4p \,\, ^3 \! P^o_{1}$ state
and 
$ 4s 4p \,\, ^3 \! P^o_{2}$ state).
%
The results of these calculations for the $ 4s 4p \,\, ^3 \! P^o_{2}$ state
are also represented by (magenta online) curves with crosses in
Figure~\ref{figure.4s4pj2-ol1}. 


\subsection{Liu~{\textit{et al}}'s approach}
\label{section.MCDHF-SDT-SP-Liu}

 {Other computation strategies have been attempted and it is worthwhile to test their coherence.  
  Liu~{\textit{et al}}~\cite{Liu:2006} focused on the spin-forbidden transition $4s^2 \; ^1S_0 - 4s4p \; ^3 \! P^o_1$ and the hyperfine-induced transition $4s^2 \; ^1S_0 -  4s4p \; ^3 \! P^o_0$ for ions between Z = 30 (Zn) and Z~=~47 (Ag).
These authors considered the following active set sequence}
 \beq
\text{AS1} &=& \{4s, 4p, 4d, 4f\}, \eol
\text{AS2} &=& \text{AS1} + \{5s, 5p, 5d, 5f, 5g\}, \eol
\text{AS3} &=& \text{AS2} + \{6s, 6p, 6d, 6f, 6g\}, \eol
\text{AS4} &=& \text{AS3} + \{7s, 7p, 7d, 7f, 7g\}, \eol
\text{AS5} &=& \text{AS4} + \{8s, 8p, 8d, 8f, 8g\}.
\eeqn{eq:LiuAS}
Their electron correlation model took into account the VV correlation,  CV correlation {through excitations of maximum one core electron from the $3d$, $3p$ and $3s$ subshells}, as well as spin-polarization {(SP)} effects by including CSFs of the forms  $1s^22s^22p^63s(ns)3p^63d^{10}$, $1s^22s(ns)2p^63s^23p^63d^{10}$ and $1s(ns)2s^22p^63s^23p^63d^{10}$. {CC correlation was systematically neglected.}
The $A$ and $B$ values that they obtained for the $4s4p\,\,^3 \! P^o_{1}$ level of $^{67}$Zn are respectively $A=20.21$~mK and $B=-0.7539$~mK,
to be compared with the two experimental results
$A=20.317(7)$~mK ($609.086(2)$~MHz)
and
$B=-0.6265(3)$~mK ($-18.782(8)$~MHz)
from
Byron~\textit{et~al}~\cite{Byron:1964}. 
The corresponding results are denoted as MCDHF-SrDT-SP-Liu. 

\subsection{Wave functions optimised  for isotope shifts}
\label{WFIS}

%
\begin{table*}[htbp!]
\caption{\small{$A$ (MHz), {$B/Q$~(MHz/b)}, and $Q$ ({b}) values {calculated with method M1  (see section~\ref{WFIS})}, as functions of the increasing active space for the $4s4p \, ^{3} \! P^{o}_{1}$ and $4s4p \, ^{3} \! P^{o}_{2}$ states in $^{67}$Zn~\textsc{i}, $I^{\pi}=5/2^-$ and $\mu_{\text{expt}}=0.875479(9)\,\mu_N$. The $Q$-values are extracted from the relation $Q=B_{\text{expt}}/{(B/Q)}$, where the experimental values are $B_{\text{expt}}(^{3} \! P^{o}_{1})=-18.782(8)^\text{a}$\,MHz and $B_{\text{expt}}(^{3} \! P^{o}_{2})= 35.806(5)^\text{b}$\,MHz.}} 
\begin{center}
\resizebox{\textwidth}{!}{
 \begin{tabular}{l r c l l l c r c l c c}
\hline
\hline
                            & \mc{5}{c}{$4s4p \, ^{3} \! P^{o}_{1}$}                                                       &~~~ & \mc{5}{c}{$4s4p \, ^{3} \! P^{o}_{2}$}                                            \\
\cline{2-6} \cline{8-12}
Active space         & $N_{\text{CSFs}}$ & & $A$~(MHz)             & \multicolumn{1}{c}{{~$B/Q$~(MHz/b)~}}   & $Q$~(b) & & $N_{\text{CSFs}}$ & & $A$~(MHz) & {~$B/Q$~(MHz/b)~}  & $Q$~(b) \\
\hline
\mc{12}{c}{MCDHF-SrDT-SP (VV+CV)}                                                                                                                                                                                          \\
$5s5p4d4f$         & $1\,592$              & & $558.02$                & $-131.036$ & $0.1433$  & & $2\,122$               & & $483.71$     & $254.975$  & $0.1404$ \\
$6s6p5d5f5g$     & $11\,932$            & & $590.45$                & $-146.084$ & $0.1286$  & & $16\,961$             & & $507.74$     & $280.708$  & $0.1276$ \\
$7s7p6d6f6g6h$ & $48\,574$            & & $610.80$                & $-150.997$ & $0.1244$  & & $71\,610$             & & $529.87$     & $290.233$  & $0.1234$ \\
$8s8p7d7f7g7h$ & $128\,264$          & & $613.17$                & $-152.617$ & $0.1231$  & & $191\,495$           & & $532.46$     & $292.535$  & $0.1220$ \\
$9s9p8d8f8g8h$ & $267\,998$          & & $617.02$                & $-154.391$ & $0.1217$  & & $402\,586$           & & $536.97$     & $296.441$  & $0.1208$ \\
$10s10p9d9f9g9h$ & $484\,772$      & & $618.47$                & $-154.071$ & $0.1219$  & & $730\,853$           & & $537.48$     & $294.773$  & $0.1215$ \\
Liu~{\textit{et al}}~\cite{Liu:2006} &   & & $605.9$                  & $-150.7$     & $0.1247$  & &                              & &                     &                    &                 \\
Expt.                   &                              & & $609.086(2)^\text{a}$ &              &                   & &                              & & $531.987(5)^\text{b}$      &                 \\
\hline
\hline
\end{tabular}
}
\end{center}
\vspace{-0.5cm}
\begin{flushleft}
\small{$^\text{a}$Byron~\textit{et~al}~\cite{Byron:1964}.} \\
\small{$^\text{b}$Lurio~\cite{Lurio:1962}.}
\end{flushleft}
\label{table_A_EFG_Q1}
\end{table*}
%

{Relativistic MCDHF wave functions have been recently optimised for estimating the electronic isotope shift parameters of $4s^2 \; ^1S_0 - 4s4p \; ^3 \! P_1^o$ and $4s4p \; ^3 \! P_2^o - 4s5s \; ^3 \! S_1 $ by Filippin~\textit{et~al}~\cite{Filetal:2017a}. Oppositely to hyperfine parameters, a reliable calculation of transition isotope shifts requires a {correct balance of electron correlation effects}
between the levels involved. These authors attempted three different strategies, systematically omitting core-core correlation in the variational process of orbital optimisation. It is indeed well-known that CC correlation effects are better balanced with the use of a common orbital basis for describing both states involved in a given transition. Neglecting CC enables to get separate orbital basis sets to allow orbital relaxation. It is interesting to investigate the hyperfine constants calculated with these computational strategies. In the present work, we estimate the hyperfine structure parameters using three approaches labelled hereafter M1, M2 and M3.} 


{The first approach (M1) was inspired by the strategy of
Liu~{\textit{et al}}~\cite{Liu:2006}, also omitting core-core correlation}.
Single (S) and double (D) substitutions were performed on a single-reference (SR) set. These SD-SR substitutions take into account valence-valence (VV) and core-valence (CV) correlations. A VV correlation model only allows SD substitutions from valence orbitals, while the VV+CV correlation model considers SrDT substitutions (single plus restricted double and triple) from core and valence orbitals, limiting the substitutions to a maximum of one hole in the core. \textit{Separate} orbital basis sets were optimised for the two studied states {$^3 \! P^o_{1,2}$}. {One difference with respect to the procedure reported in~\cite{Filetal:2017a} for isotope shift parameters is that the $1s$ shell is opened in the present work to include the spin-polarisation effects that are relevant for hyperfine structure calculations. The procedure can be {outlined}
as follows:}

(1) Perform a calculation using a set consisting of CSFs with two forms: [Ar]$3d^{9}nln'l'n''l'' \, J^\Pi$ and [Ne]$3s^23p^53d^{10}nln'l'n''l'' \, J^\Pi$ with $n,n',n''=4$ and $l,l',l''=s,p,d,f$,
plus $5s$ and $5p$. These CSFs account for a fair amount of the VV correlation, and for CV correlations between the $3p$ and $3d$ core orbitals and the $5s$, $5p$ and $n=4$ valence orbitals.
Add spin polarisation (SP) by including the following CSFs: \\
$1s^22s^22p^63s3p^63d^{10}4s4p5s \, J^\Pi$, \\
$1s^22s2p^63s^23p^63d^{10}4s4p5s \, J^\Pi$, \\ 
$1s2s^22p^63s^23p^63d^{10}4s4p5s \, J^\Pi$.

(2) Keep the orbitals fixed from step (1), and optimise an orbital basis, layer by layer, up to an active space equal to $10s10p9d9f9g9h$, described by CSFs with the $J^\Pi$ symmetry of the state. These CSFs are obtained by SrDT-SP substitutions as in step (1) (at most one substitution from the $1s^22s^22p^63s^23p^63d^{10}$ core~\footnote{The original notation adopted in~\cite{Filetal:2017a} was (SrDT-SS) where the `SS' stands for ``single $s$ substitutions'' describing spin-polarisation (SP) of the core $s$ subshells.}).

{The corresponding results are presented in \tablename{~\ref{table_A_EFG_Q1}.
The MCDHF-SrDT-SP-Liu active space expansion used in~\cite{Liu:2006}} {optimised simultaneously the $^3 \! P^o_1$ and $^3 \! P^o_2$ levels}. Therefore the $A$, {$B/Q$} and $Q$ results obtained {in the present work} slightly differ from {those reported in~\cite{Liu:2006}} for  $J=1$.} 
%


%
\begin{table*}[htbp!]
\caption{\small{MR configurations for the $4s4p \, ^{3} \! P^{o}_{1}$ and $4s4p \, ^{3} \! P^{o}_{2}$ states in $^{67}$Zn~\textsc{i}. The MR-cutoff value, $\varepsilon_{\text{MR}}$, determines the set of CSFs in the MR space. $N_{\text{CSFs}}$ is the number of CSFs describing each MR space.}}
\begin{center}
\begin{tabular}{l c l c l l c r}
\hline
\hline
\vspace{0.1cm}
State                & & $\varepsilon_{\text{MR}}$ & & \mc{2}{l}{MR configurations}                                                                                           & & $N_{\text{CSFs}}$ \\
\hline
$4s4p \, ^{3} \! P^{o}_{1}$ & & 0.01                      & & [Ar]$3d^{10}\{4s4p,4p4d\}$, & [Ar]$3d^{9}\{4s4p4d,4s4d4f,4p^3,4p^24f,4s^24p\}$ & & 31                         \\
\vspace{0.1cm}
$4s4p \, ^{3} \! P^{o}_{2}$ & & 0.01                      & & [Ar]$3d^{10}\{4s4p,4p4d\}$, & [Ar]$3d^{9}\{4s4p4d,4s4d4f,4p^3,4p^24f,4s^24p\}$ & & 31                         \\
\hline
\hline
\end{tabular}
\end{center}
\label{table_MR_composition}
\end{table*}
%

%
\begin{table*}[htp!]
\caption{\small{$A$ (MHz), {$B/Q$~(MHz/b)}, and $Q$ ({b}) values {calculated with methods M2 (upper part) and M3 (lower part) - see section~\ref{WFIS})}, as functions of the increasing active space for the $4s4p \, ^{3} \! P^{o}_{1}$ and $4s4p \, ^{3} \! P^{o}_{2}$ states in $^{67}$Zn~\textsc{i}. $I^{\pi}=5/2^-$ and $\mu_{\text{expt}}=0.875479(9)\,\mu_N$. The $Q$-values are extracted from the relation $Q=B_{\text{expt}}/\text{EFG}$, where the experimental values are $B_{\text{expt}}(^{3} \! P^{o}_{1})=-18.782(8)^\text{a}$\,MHz and $B_{\text{expt}}(^{3} \! P^{o}_{2})= 35.806(5)^\text{b}$\,MHz.}}
\begin{center}
\resizebox{\textwidth}{!}{
\begin{tabular}{l r c l l l c r c l l l}
\hline
\hline
                                  & \mc{5}{c}{$4s4p \, ^{3} \! P^{o}_{1}$}                                            & & \mc{5}{c}{$4s4p \, ^{3} \! P^{o}_{2}$}                                    \\
\cline{2-6} \cline{8-12}
Active space               & $N_{\text{CSFs}}$ & & $A$~(MHz)  & {~$B/Q$~(MHz/b)~}   & $Q$~(b) & & $N_{\text{CSFs}}$ & & $A$~(MHz) & {~$B/Q$~(MHz/b)~}   & $Q$~(b) \\
\hline
$4s4p3d$~({DHF})                        & $2$                 & & $475.27$   & $-100.437$ & $0.1870$   & & $1$                       & & $419.98$   & $192.166$  & $0.1863$ \\
\mc{12}{c}{{MCDHF-SrD-SR~(VV+CV)}} \\
$5s5p4d4f$                     & $1\,454$         & & $554.21$   & $-129.614$ & $0.1449$   & & $2\,108$               & & $483.74$   & $253.875$  & $0.1410$ \\
$6s6p5d5f5g$                 & $5\,857$         & & $590.45$   & $-145.650$ & $0.1290$   & & $5\,790$               & & $509.32$   & $280.362$  & $0.1277$ \\
$7s7p6d6f6g6h$             & $14\,381$       & & $617.54$   & $-152.198$ & $0.1234$   & & $14\,467$             & & $534.86$   & $292.975$  & $0.1222$ \\
$8s8p7d7f7g7h$             & $27\,052$       & & $627.21$   & $-157.702$ & $0.1191$   & & $27\,426$             & & $542.93$   & $303.008$  & $0.1182$ \\
$9s9p8d8f8g8h$             & $43\,870$       & & $627.36$   & $-159.448$ & $0.1178$   & & $44\,667$             & & $547.10$   & $305.735$  & $0.1171$ \\
$10s10p9d9f9g9h$         & $64\,835$       & & $631.06$   & $-159.859$ & $0.1175$   & & $66\,190$             & & $550.27$   & $306.528$  & $0.1168$ \\
$11s11p10d10f10g10h$ & $89\,947$       & & $632.63$   & $-159.987$ & $0.1174$   & & $91\,995$             & & $550.84$   & $306.521$  & $0.1168$ \\
+ $1s$ open                    & $95\,907$       & & $638.82$   & $-159.974$ & $0.1174$   & & $97\,610$             & & $556.64$   & $306.592$  & $0.1168$ \\
\mc{12}{c}{{CI-SD-SR~(VV+CV+CC)}} \\
$11s11p10d10f10g10h$ & $1\,236\,101$ & & $546.09$ & $-129.845$ & $0.1446$    & & $1\,243\,611$       & & $479.14$   & $249.303$  & $0.1436$ \\
+~t(MCHF) $9s9p8d8f8g$ &               & & $578.05$ & $-145.095$ & $0.1294$    & &                                & & $506.03$   & $279.803$  & $0.1280$ \\
                                  &                        & &                    &                    &                   & &                             & &                    &                   &                 \\   
\mc{12}{c}{Multireference calculations}  \\                                                                                                                                                                       
$5s5p4d4f$~(MR)              & $903$             & & $541.40$  & $-112.267$ & $0.1673$    & & $1\,231$               & & $414.61$   & $198.252$  & $0.1806$ \\
\mc{12}{c}{MCDHF-SrD-MR (VV+CV)}  \\ 
$6s6p5d5f5g$                 & $12\,015$       & & $595.05$  & $-148.023$ & $0.1269$    & & $16\,521$             & & $511.98$   & $284.444$  & $0.1259$ \\
$7s7p6d6f6g6h$             & $32\,172$       & & $621.07$  & $-153.354$ & $0.1225$    & & $45\,722$             & & $535.07$   & $294.919$  & $0.1214$ \\
$8s8p7d7f7g7h$             & $62\,730$       & & $631.14$  & $-159.206$ & $0.1180$    & & $90\,401$             & & $544.19$   & $306.010$  & $0.1170$ \\
$9s9p8d8f8g8h$             & $103\,689$     & & $630.28$  & $-160.573$ & $0.1170$    & & $150\,558$           & & $548.36$   & $308.353$  & $0.1161$ \\
$10s10p9d9f9g9h$         & $155\,049$     & & $633.80$  & $-160.991$ & $0.1167$    & & $226\,193$           & & $551.29$   & $309.010$  & $0.1159$ \\
$11s11p10d10f10g10h$ & $216\,810$     & & $635.32$  & $-161.006$ & $0.1167$    & & $317\,306$           & & $551.96$   & $308.985$  & $0.1159$ \\
+ $1s$ open                    & $232\,787$     & & $641.60$  & $-161.025$ & $0.1167$    & & $339\,230$           & & $557.60$   & $309.005$  & $0.1159$ \\
\mc{12}{c}{\ }  \\   
Expt.                               &                         & & $609.086(2)^\text{a}$ & &                   & &                               & & $531.987(5)^\text{b}$ & &                 \\
\hline
\hline
\end{tabular}
}
\end{center}
\vspace{-0.5cm}
\begin{flushleft}
\small{$^\text{a}$Byron~\textit{et~al}~\cite{Byron:1964}.} \\
\small{$^\text{b}$Lurio~\cite{Lurio:1962}.}
\end{flushleft}
\label{table_A_EFG_Q2}
\end{table*}
%


{The second approach  (M2) considered single and restricted double substitutions performed on a single-reference (SR) set (MCDHF-SrD-SR).} The VV correlation model only allows SD substitutions from valence orbitals, while the VV+CV correlation model considers SrD substitutions from core and valence orbitals, limiting the substitutions to a maximum of one hole in the core. {Oppositely to Filippin~\textit{et~al}~\cite{Filetal:2017a}, core correlation is included in the final step through a configuration interaction calculation based on SD-SR expansions. The corrections {for} triple excitations estimated in section~\ref{subsection.triples-corrections} (see \tablename~\ref{table.triples-correctionsAEFG}),  are added in the very final step. This computational strategy can be {outlined} by the following sequence:}

(1) Run a calculation using SR set consisting of CSF(s) of the form 
%
$2s^22p^63s^23p^63d^{10}4s4p \, J^\Pi$.

(2) Keep the orbitals fixed from step (1), and optimise an orbital basis layer by layer up to an active space equal to $11s11p10d10f10g10h$, described by CSFs with the $J^\Pi$ symmetry of the state. These CSFs are obtained by SrD-SR substitutions (at most one substitution from the $2s^22p^63s^23p^63d^{10}$ core).

(3) Perform a CI calculation on the CSFs expansion with the $J^\Pi$ symmetry of the state, describing VV, CV and CC correlation obtained by SD-SR substitutions to the orbital basis from step (2).

(4) Add a correction to the $A$ and {EFG} values from step~(3), accounting for triple (t) substitutions and obtained from a {non-relativistic} MCHF computation.

{Step (3), allowing the inclusion of core-core correlation through CI, and 
step (4), specific to hyperfine constants, were not considered in~\cite{Filetal:2017a}. The corresponding results are denoted
MCDHF-SrD-SR/CI-SD-SR+t(MCHF)
in the upper part of \tablename{~\ref{table_A_EFG_Q2}}. The final line shows}
that opening $1s$ brings a non-negligible contribution to the $A$-values, which become approximately 6 MHz larger for both states. 
{This approach is very similar to the method described in section~\ref{subsection.GRASP-4s4pOL1}.
{However}, the advantage of the M2 calculations is {in the simultaneous generation of}
both $4s4p \; ^3 \! P^o_2$ and $4s4p \; ^3 \! P^o_1$ {states},
at the Optimal Level (OL)
{of the variational functional.}
For $4s4p \; ^3 \! P^o_2$, the results are indeed very similar to the MCDHF-SD-SR-OL1+t(MCHF) values already reported in Table~\ref{table.summaryAQZn} and discussed in section~\ref{subsection.GRASP-4s4pOL1} dedicated to that level.  For these M2 results, only the  $4s4p \; ^3 \! P^o_1$ values are therefore reported in the final summary {in} Table~\ref{table.summaryAQZn}. }


{The third approach (M3)  considered SrD substitutions performed on a multi-reference (MR) set. The latter contains the CSFs that have large expansion coefficients and account for the major correlation effects. For building this MR set, a MCDHF calculation is first performed using a CSF expansion based on SrDT substitutions from the $3d$ and the occupied valence orbitals towards the $5s$, $5p$ and $n=4$ valence orbitals (maximum of one hole in the $3d$ orbital).} 

Due to limited computer resources, such an MR set would be too large for subsequent calculations. Hence, only the CSFs whose expansion coefficients are, in absolute value, larger than a given MR cutoff are kept, i.e., $\vert c_\nu \vert >\varepsilon_{\text{MR}}$. {The resulting MR sets are 
{outlined}
in \tablename{~\ref{table_MR_composition}}.} Only orbitals occupied in the single configuration DHF approximation are treated as spectroscopic, and the occupied reference orbitals are kept frozen in the subsequent calculations. 

{The M3 procedure consists in the following sequence:}

(1) Perform a calculation using an MR set consisting of CSFs with two forms: \\
$2s^22p^63s^23p^63d^{10}nln'l' \, J^\Pi$ with $n,n'=4$ and $l,l'=s,p,d,f$ + $5s$ and $5p$, and \\$2s^22p^63s^23p^63d^9nln'l'n''l'' \, J^\Pi$ with $n,n',n''=4$ and $l,l',l''=s,p,d,f$ + $5s$ and $5p$. These CSFs account for a fair amount of the VV correlation, and for CV correlations between the $3d$ core orbital and the $5s$, $5p$ and $n=4$ valence orbitals.
{Keep in the MR set the CSF whose expansion coefficients are, in absolute value, larger than $\epsilon_{\text{MR}} = 0.01$.}

(2) Keep the orbitals fixed from step (1), and optimise an orbital basis layer by layer up to an active space equal to $11s11p10d10f10g10h$, described by CSFs with the $J^\Pi$ symmetry of the state. These CSFs are obtained by SrD-MR substitutions (at most one substitution from the $2s^22p^63s^23p^63d^{10}$ core).
%
As observed for M2, spin-polarisation of the $1s$ shell is not negligible.
{The results of these calculations are presented in the lower part of 
\tablename{~\ref{table_A_EFG_Q2}},
and
labeled MCDHF-SrD-MR 
in \tablename{~\ref{table.summaryAQZn}}.


\begin{widetext}
%
\section{Evaluation of the nuclear quadrupole moment of $Q(^{67}{\rm Zn})$}
\label{section.Q-4s4p}

\begin{table*}[htbp!]
\caption{Summary of the $A$ (MHz), {~$B/Q$~(MHz/b)~}, and $Q$ (b) values of $^{67}$Zn.}
\label{table.summaryAQZn}  
\begin{tabular}{l c l c l c l c l}
\hline
\hline
 State & & \multicolumn{1}{c}{$A$ (MHz)} & & \multicolumn{1}{l}{{~$B/Q$~(MHz/b)~}} & & \multicolumn{1}{c}{$Q$ (b)} & & Method \\
\colrule 
&&&&&&&& \\
  $4s4p\,\,^3 \! P^o_{1}$ & &      634.802        & & $-$165.058 
                 & & 0.113790                                 & & MCHF-SD(T) \\
                                    & & 605.9            & & $-$150.7                                 & & 0.1247                                     & & MCDHF-SrDT-SP-Liu \\
                                    & & 618.47          & & $-$154.071                             & & 0.121905                                 & & MCDHF-SrDT-SP \\
                                    & & 641.60          & & $-$161.025                             & & 0.116640                                 & & MCDHF-SrD-MR \\
                                    & & 578.05          & & $-$145.095                             & & 0.129446                                 & &
{MCDHF-SrD-SR/CI-SD-SR+t(MCHF)} \\
                                    & & 577.886        & & $-$142.579                             & & 0.131730                                 & & MCDHF-SD-SR-OL4+t(MCHF) \\
                                    & & 609.086(2)$^\text{a}$  & &                               & &                                                 & & Expt. \\
&&&&&&&& \\
$4s4p\,\,^3 \! P^o_{2}$ & & 560.310        & & \;\, 317.254                            & & 0.112862                                 & & MCHF-SD(T) \\
                                    & & 537.48          & & \;\, 294.773                            & & 0.121470                                 & & MCDHF-SrDT-SP \\
                                    & & 557.60          & & \;\, 309.005                            & & 0.115875                                 & & MCDHF-SrD-MR \\
                                    & & 509.861        & & \;\, 281.799                            & & 0.127062                                 & & MCDHF-SD-SR-OL1+t(MCHF) \\
                                    & & 513.200        & & \;\, 271.989                            & & 0.131645                                 & & MCDHF-SD-SR-OL4+t(MCHF) \\
                                    & & 531.987$^\text{b}$  & &                                    & &                                                 & & Expt. \\
\hline
\hline
\end{tabular}
\begin{flushleft}
\small{$^\text{a}$Byron~\textit{et~al}~\cite{Byron:1964}.} \\
\small{$^\text{b}$Lurio~\cite{Lurio:1962}.}
\end{flushleft}
\end{table*}
%
%
%
%
We {report} in the present section eleven calculated $Q$ {(and $A$) values},
obtained with the following approaches: 
%
\begin{itemize}
\item MCHF-SD/CI-SDT+DHF/HF correction ($^3 \! P^o_{1,2}$ levels), {under label MCHF-SD(T), - see sections~\ref{section.MCHF.4s4p} and \ref{section.DHF-HF-correction},} 
\item
MCDHF-SrDT-SP-Liu: from Liu~\textit{et~al}~\cite{Liu:2006} (only $^3 \! P^o_1$ level), {- see section~\ref{section.MCDHF-SDT-SP-Liu}}, 
\item
MCDHF-SrDT-SP: calculation based on Liu~\textit{\textit{et~al}}'s strategy ($^3 \! P^o_{1,2}$ levels) { - see method M1 in section~\ref{WFIS}}, 
%
\item
MCDHF-SrD-SR/CI-SD-SR+t(MCHF):
single-reference + MCHF triples correction ($^3 \! P^o_{1}$ level~\footnote{The corresponding $^3 \! P_{2}\,\!^o$ results are not reported in Table~\ref{table.summaryAQZn} since the strategy is similar to the MCDHF-SD-OL1+t(MCHF) method (see section~\ref{subsection.GRASP-4s4pOL1}), with consistent results.}) { - see method M2 in section~\ref{WFIS}}, 
\item
MCDHF-SrD-MR: multi-reference ($^3 \! P^o_{1,2}$ levels),
{ - see method M3 in section~\ref{WFIS}},  
\item
MCDHF-SD-OL1+t(MCHF): OL1 ($J=2$) single reference + MCHF triples correction
(only $^3 \! P^o_2$ level) { - see section~\ref{subsection.GRASP-4s4pOL1}}, 
\item
MCDHF-SD-OL4+t(MCHF): OL4 ($J=0,1,1,2$) single reference + MCHF triples correction ($^3 \! P^o_{1,2}$ levels)
 { - see section~\ref{subsection.GRASP-4s4pOL4}}, 
\end{itemize}
%
where the shorthand notations above represent the following computational
methods: \\
%
\begin{tabular}{ll}
MCHF & MultiConfiguration Hartree-Fock (non-relativistic) \\
CI-SDT & Configuration Interaction Hartree-Fock (non-relativistic) \\
DHF/HF & multiplicative relativistic correction described in
         section~\ref{section.Non-relativistic} \\
MCDHF & MultiConfiguration Dirac-Hartree-Fock (relativistic)  \\
SD & Single and Double substitutions in the SCF process   \\
SrD & Single and restricted Double substitutions in the SCF process   \\
t(MCHF) & additive correction for triple substitutions estimated from MCHF calculation  {(see section~\ref{subsection.triples-corrections}).} \\
SP & Spin Polarisation (method described in
   Liu~\textit{et~al}~\cite{Liu:2006}) \\
SR & Single-Reference  \\
MR & Multi-Reference  \\
OL1 & Optimal Level calculation with optimisation on one level {(J=2)}  \\
OL4 & Optimal Level calculation with optimisation on four levels  \\
\end{tabular}
\ \\
\end{widetext}

We {adopted} a convention used by chemists, where T in parentheses
(T) implies that 
triple substitutions are included in a post-SCF approach
(M\o{}ller-Plesset or CI or another method).
In our notation
{t(MCHF)}
means an additive correction for triple substitutions
evaluated {with} the ATSP code~\cite{ATSP2k:2007}.
%
The calculated {EFGs} were combined with the measured $B$ values for the
$ 4s 4p \,\, ^3 \! P^o_{1} $ state~\cite{Byron:1964}, and for the
$ 4s 4p \,\, ^3 \! P^o_{2} $ state~\cite{Lurio:1962}
of the neutral Zn atom,
to yield eleven calculated values of 
$Q$($^{67}$Zn),
presented in the fourth column of \tablename{~\ref{table.summaryAQZn}}.

Although these eleven values do not represent the sample
in the statistical sense, the scatter of the values gives us an information
about {the} dependence of the calculated values of {EFG} on the choice of
the method
of calculation, and provides a basis for an estimate of the error bar for
the determination of the quadrupole moment $Q(^{67}{\rm Zn})$.
We assumed that the error bar should at least overlap with all eleven results.
For computing the final value of $Q(^{67}{\rm Zn})$ one might consider 
taking the average of the results of the eleven calculations
($Q = 0.1208$~{b}),
or the median value thereof
($Q = 0.1223$~{b}); both methods yield very close results,
the difference being negligible compared to the error bar resulting
from the arguments presented above.
Assuming the above procedures and estimates, we arrived at
$Q(^{67}{\rm Zn}) = 0.12 \pm 0.01$~{b},
obtained from
the $4s 4p \,\, ^3 \! P^o_{1,2}$ states of zinc.
The relative error bar (8~\%) is of the same order
as the error bar (10~\%) associated with the previous standard value,
$Q(^{67}{\rm Zn}) = 0.150(0.015)$~{b},
quoted by Stone~\cite{NJStone:2016}
and by Pyykk\"o~\cite{Pyykko:08bb}, 
and based on measurements performed
by Laulainen and McDermott~\cite{LaulainenMcDermott:1969},
but the $Q(^{67}{\rm Zn})$ value itself is now downshifted by 20~\%.
On the other hand, our value is in very good agreement with 
$Q(^{67}{\rm Zn})$ = 0.125(5)~{b}, of
Haas~{\textit{et al}}~\cite{Haas:2017},
who used a hybrid Density Functional Theory approach.

An inspection of the results presented in the 
\tablename{~\ref{table.summaryAQZn}}
leads to the conclusions, that the
multi-reference MCDHF-SrD-MR calculations overshoot the values
of the magnetic dipole hyperfine constant $A$
by about $20-25$~MHz,
while single reference MCDHF-SD-SR-OL1+t(MCHF) and MCDHF-SD-SR-OL4+t(MCHF)
results for $A$ are too small by nearly the same amount.
The best agreement with the experimental $A$ value was obtained
in the calculation of Liu~\textit{et~al}~\cite{Liu:2006},
which is understandable, since, as mentioned in
section~\ref{section.MCDHF-SDT-SP-Liu},
the main objective of Liu~\textit{et~al}
was the magnetic dipole hyperfine structure,
therefore they carefully treated the spin-polarization effects.
Incidentally, the nuclear quadrupole moment $Q(^{67}{\rm Zn})$
calculated from their {EFG} value is in fact 
quite close to the median value $Q = 0.1223$~{b}, mentioned above.
The abovementioned discrepancies may be assumed as another tool to estimate
the error bar for determination of $Q$.
The error bar estimate from $A$ is of the order of 4~\%, smaller than
that obtained from the sample of eleven $Q$ values.
We  assumed the larger of the two error bar estimates,
and finally we propose
\begin{equation}                                  
Q(^{67}{\rm Zn}) = 0.122(10)~{\mbox{b}} \; . 
\end{equation}   
This value has been utilized to extract electric quadrupole moments
of odd-$A$ nuclei in the range $A$~=~63--79
across the isotopic chain of zinc,
following the measurements of electromagnetic moments
by Wraith~\textit{\textit{et~al}}~\cite{Wraith:PLB:2017}.

\section{Conclusions}
\label{section.Conclusions}

The calculations of hyperfine shifts are inherently inaccurate
(or accurate to a few percent). We do have computational tools to estimate
accuracy of expectation
values~\cite{BieronAu2009,BieronBeF1999,Bieron:e-N:2015},
but they are more expensive computationally
than the calculations of expectation values themselves. Therefore we rarely
{\sl compute} accuracy, because normally we compute the expectation values
themselves at the limits of our computing resources, and this does not leave
enough resources for computing accuracy. Then we {\sl estimate} the accuracy.
Estimating the accuracy of a single calculation of an {EFG} for a single
level is in fact not much more than guesswork.
If magnetic dipole hyperfine coupling constant
$A$ is known (i.e.~measured $A_{\text{expt}}$ exists)
then accuracy of {EFG} is sometimes assumed from the difference
$A_{\text{expt}} - A_{\text{calc}}$. 
Another method is to carry out calculations with several different
methods and evaluate the accuracy from differences between the results
obtained with those methods.
In the present paper the latter approach yields the larger error bar.
The optimal method would be to carry out measurements and
calculations for several levels.
From this point of view having hyperfine structure data for several levels 
would give us a benefit of more tools to estimate accuracy.
Combined with measured values of $A$ and $B$
for these levels, we would obtain a {\sl statistical} sample for both $A$
and {EFG}.
It is not exactly statistical because calculations 
are in principle not fully independent, 
but several levels is still better than one or two levels.

\begin{acknowledgments}

\noindent
The large-scale calculations were carried out with the supercomputer Deszno
purchased thanks to the financial support of the European Regional
Development Fund in the framework of the Polish Innovation
Economy Operational Program (contract no. POIG.02.01.00-12-023/08).
Computational resources have also been provided by the Shared ICT
Services Centre, Universit{\'e} libre de Bruxelles and by the Consortium
des {\'E}quipements de Calcul Intensif (C{\'E}CI), funded by the Fonds de la
Recherche Scientifique de Belgique (F.R.S.-FNRS) under Grant No.~2.5020.11.
MG was supported by the Belgian F.R.S.-FNRS Fonds de la Recherche Scientifique
(CDR J.0047.16) and FWO \& FNRS Excellence of Science Programme (EOS-O022818F).
This work is also supported by the Swedish Research Council
under contract 2015-04842.
\end{acknowledgments}

\bibliography{xet}

\end{document}